\documentclass[twocolumn,showpacs,aps,amsmath,amssymb]{revtex4}
\begin{document}
\title{\bf Dark energy, Dirac's scalar field and the cosmological constant problem}
\author{O.V. Babourova}\email{baburova@orc.ru}
\author{B.N. Frolov}\email{frolovbn@orc.ru}
\affiliation{Moscow Pedagogical State  University,Faculty of Physics and 	Information Technologies, Malaya Pirogovskaya ul. 29, Moscow 119992, 
\\ Russian Federation}

\begin{abstract}
The solutions of the field equations of the conformal theory of gravitation with Dirac scalar field in Weyl--Cartan spacetime in very early universe are obtained. In this theory dark energy (describing by an effective cosmological constant) is a function of the  Dirac scalar field $\beta$. These solutions describó  the exponential decreasing of $\beta$ at the first exponential stage of inflation. One of these solutions has a limit to a constant value of the dark energy at large time that can give a way to solving the fundamental cosmological constant problem as a consequence of the fields dynamics in the early universe. 
\end{abstract}
\maketitle

On the basis of the observational data, it is accepted in modern cosmology that the dark energy (described by the cosmological constant) is of dominant importance in dynamics of the universe. In this connection the major unsolved problem of modern fundamental physics is very large difference of around 120 orders of magnitude between a very small value of Einstein's cosmological constant $\Lambda$, which can be estimated on the basis of modern observations in cosmology, and theoretical calculation in quantum field theory of quantum fluctuation contributions to the vacuum energy \cite{Wein}.

We use Poincare--Weyl gauge theory of gravitation that has been developed in \cite{BFZ1}, \cite{BFZ2}. This theory is invariant both concerning the Poincare subgroup and the Weyl subgroup  --  extensions and compressions (dilatations) of spacetime. In this theory the additional scalar field $\beta(x)$ is introduced as an essential geometrical component of the metric tensor. This field is the same as the scalar field introduced by Dirac \cite{Dir}. In this theory spacetime has the geometrical structure of Weyl--Cartan space with curvature, torsion $T^a{}_{\mu\nu}$ and nonmetricity $Q^{\alpha\beta}{}_\mu$ of the Weyl type: $Q^{\alpha\beta}{}_\mu =\frac14 g^{\alpha\beta}Q_\mu$, where $Q_\mu = Q^\nu{}_{\nu\mu}$ is a Weyl vector.

We use the tetrad formalism, in which the Lagrange density of the theory is as follows,
\begin{eqnarray*}
&\mathcal{L} = \mathcal{L}_G + \mathcal{L}_m + \frac12\sqrt{-g}\Lambda^\mu{}_{ab}\left(Q^{ab}{}_\mu -\frac14 g^{ab}Q_\mu\right)\,,& \nonumber\\ 
&\Lambda^\mu{}_{ab}g^{ab} = 0\,,&
\end{eqnarray*}
where $\Lambda^\mu{}_{ab}$ are Lagrange multipliers, $\mathcal{L}_m$ is a Lagrange density of matter. The  Lagrange density of gravitational field we present as follows 
($c = 1$), 
\begin{eqnarray} 
&&\mathcal{L}_G= \sqrt{-g}(f_0\beta^2 R + L_{R^2}+\beta^2 L_{T^2} +\beta^2 L_{Q^2}  \nonumber\\
&&+ \beta^2 L_{TQ} + L_\beta)\,, \quad f_0=\frac{1}{2\varkappa}\,,\quad 
\varkappa = 8\pi G\,. \label{eq:LGG}
\end{eqnarray}
Here $L_{R^2}= (\lambda/4)V^{\mu\nu}V_{\mu\nu}$ \cite{CQG}, where $V_{\mu\nu} = R^a{}_{a\mu\nu} = \partial_{[\mu} Q_{\nu ]}$ is a Weyl's segmental curvature tensor. The Lagrange density (\ref{eq:LGG}) includes also Lagrangians quadratic in torsion, $L_{T^2}$,  Lagrangians $L_{QT}$ containing curvature--torsion interactions, and a proper Lagrangian of the scalar field
\begin{eqnarray*}
&&L_{\beta}  = l_1g^{\mu\nu}\partial_\mu\beta\partial_\nu\beta + l_2\beta\partial_\mu\beta g^{\mu\sigma} T_\sigma \nonumber\\
&&+ l_3\beta\partial_\mu\beta g^{\mu\sigma}Q_\sigma 
+ l_4\beta \partial_\mu\beta Q^{\mu\sigma}{}_{\sigma} + \Lambda \beta^4\,.
\end{eqnarray*}

Variational equations of the field in the Weyl--Cartan spacetime have been derived by variation of the full Lagrangian density of the theory. The independent variables are tetrads $h^a{}_\mu$, a nonholonomic connection $\Gamma^a{}_{b\mu}$, the scalar field $\beta$ and the Lagrange multipliers. Variational equations are the equations of the gravitational field of the conformal theory of gravity in the Weyl--Cartan spacetime, see them in \cite{BFK}. These equations are now  investigated with a view to obtaining and solving the equations for the scale factor $a(t)$ and the scalar Dirac field $\beta$ at the very early stage of evolution of universe, when a matter density has been very small, $\mathcal{L}_m\approx 0$.

In  homogeneous and isotropic spacetime the conditions,  $T^a{}_{\mu\nu} = -\frac{2}{3}h^a{}_{[\mu}T_{\nu]}$ \cite{Tsamp} and $V_{\mu\nu} = 0$ \cite{CQG}, are valid. By means of these conditions we can express from the $\Gamma$-equation a torsion trace and Weyl vector through the Dirac scalar field $\beta$ \cite{BFK}:
\begin{equation}
T_\mu = \chi_T\partial_\mu\ln\beta\,, \qquad
Q_\mu = \chi_Q\partial_\mu\ln\beta\,.
\label{eq:TQb}
\end{equation}
The coefficients in (\ref{eq:TQb}) are expressed through couple constants of the Lagrangian density (\ref{eq:LGG}).

Then the $h-$equation together with the $\beta-$equation, taking into account (\ref{eq:TQb}), are as follows, 
\begin{eqnarray}
&\stackrel{R}{R}_{\alpha\beta} - \frac{1}{2}g_{\alpha\beta}\stackrel{R}{R} - 2\stackrel{R}{\nabla_\alpha}\stackrel{R}{\nabla_\beta}\ln\beta +
2B_1 \partial_\alpha \ln\beta \partial_\beta\ln\beta  & \nonumber \\
&+ g_{\alpha\beta} g^{\mu\nu}(2\stackrel{R}{\nabla_\mu}\stackrel{R}{\nabla_\nu}\ln\beta + B_2\partial_\mu \ln\beta \partial_\nu\ln\beta )&\nonumber \\
&- g_{\alpha\beta}\Lambda \beta^2 = \varkappa\, t^{(m)}_{\alpha\beta}\,,& 
\label{eq:RR}\\
&Ag^{\mu\nu}\stackrel{R}{\nabla_\mu}\stackrel{R}{\nabla_\nu}\ln\beta + B g^{\mu\nu}\partial_\mu \ln\beta \partial_\nu\ln\beta = 0\,,&
\label{eq:BB}
\end{eqnarray}
where "R" under the quantities means that these quantities relay to Riemann spacetime. Here the constants $A$, $B$, $B_1$ and $B_2$ are expressed through the parameters of the Lagrangian (\ref{eq:LGG}).

Now we consider the spatially flat Friedman--Robertson--Walker (FRW) metric 
\begin{equation}
ds^2 = dt^2 - a^2(t)(dx^2 + dy^2 + dz^2)\,.
\label{eq:Frid}
\end{equation}
Taking into account $t^{(m)}_{\alpha\beta}\approx 0$, we obtain from (\ref{eq:RR}), (\ref{eq:BB}) the following system of equations, 
\begin{eqnarray}
(0,\,0):& 3\frac{\dot{a}^2}{a^2} + 6\frac{\dot{a}}{a}\frac{\dot{\beta}}{\beta} 
+ 3B_3 \left (\frac{\dot{\beta}}{\beta}\right )^2 = \Lambda{\beta}^2\,&\label{eq:ur00}\\
(1,\,1):& 2\frac{\ddot{a}}{a} + 2\frac{\ddot{\beta}}{\beta} +4\frac{\dot{a}}{a}\frac{\dot{\beta}}{\beta} + \left (\frac{\dot a}{a}\right )^2 \nonumber\\
&+ (B_2 - 2) \left (\frac{\dot \beta}{\beta}\right )^2 = \Lambda{\beta}^2\,,& \label{eq:ur33}\\
\beta\;\;:& A\left(\frac{\ddot{\beta}}{\beta} + 3\frac{\dot{a}}{a}\frac{\dot{\beta}}{\beta}\right ) 
+ (B - A) \left (\frac{\dot \beta}{\beta}\right )^2 = 0\,.& \label{eq:ur22}
\end{eqnarray}
Here $B_3 = \frac{1}{3}(2B_1 + B_2)$, and the components (2,\,2) and (3,\,3) are equal to the component (1,\,1). 

The system of equations (\ref{eq:ur00})--(\ref{eq:ur22}) is inconsistent, because we have three equations for two unknown functions $a(t)$ and $\beta (t)$. 

Let us put in this system 
\begin{equation}
B_1 = B_2 = B_3 = 1\,, \label{eq:Bi}
\end{equation}
and also $u= \ln a$, $v = \ln \beta$. Then substract Eq. (\ref{eq:ur00}) from Eq. (\ref{eq:ur22}). As a result we obtain the following system of equations, 
\begin{eqnarray}
&& (\dot u)^2 + 2\dot{u}\dot{v} + (\dot v)^2= \frac{\Lambda}{3}e^{2v}\,\label{eq:ur0}\\
&& \ddot{u} + \ddot{v} - \dot{u}\dot{v} - (\dot v)^2= 0\,, \label{eq:ur3}\\
&& \ddot{v} + 3\dot{u}\dot{v} + \frac{B}{A } (\dot v)^2 = 0\,. \label{eq:ur2}
\end{eqnarray}

Eq. (\ref{eq:ur0}) is equivalent to the equation, 
\begin{equation}
\dot u + \dot{v} = \pm \lambda e^{v}\,,\quad \lambda = \sqrt{\frac{\Lambda}{3}}\,. \label{eq:dudv}
\end{equation}
It is easy to check that Eq. (\ref{eq:ur3}) is fulfilled identically as a consequence of Eq. (\ref{eq:dudv}). Therefore 
we have only 2 equations (\ref{eq:ur2}) and (\ref{eq:dudv}) for 2 unknown functions $a(t)$, $\beta (t)$, and this system of equations is consistent. In what follows we choose the sign "--" in Eq. (\ref{eq:dudv}). 

Let us find $\dot u$ from  Eq. (\ref{eq:dudv}) and put it in Eq.(\ref{eq:ur2}). We obtain the equation, 
\begin{equation}
\ddot v - 3\lambda e^{v}\dot{v} + \omega (\dot v)^2 = 0  \,, \quad \omega = \frac{B}{A} - 3 \,. \label{eq:ddv}
\end{equation}
The first integral of this equation is the following, 
\begin{equation}
\dot v = \lambda_1 \beta^{-\omega} - \frac{3\lambda}{1+\omega}\beta^{\omega}\,, \label{eq:bdot}
\end{equation}
where $\lambda_1$ is a constant of integration. 

Let us consider the simplist case, when 
\begin{equation}
\omega = 0\,, \qquad B = 3A \,.\label{eq:B3A}
\end{equation}
In this case the second integration yields
\begin{equation}
\beta = \beta_0 e^{-\sigma t}\,, \qquad \sigma = -\lambda_1 + 3\lambda \,, \label{eq:Bt}
\end{equation}
where $\beta_0$ is the second constant of integration. 

Then from Eq. (\ref{eq:dudv}) we find for the scale factor,
\begin{equation}
a = a_0 e^{\sigma t}e^{(e^{-\sigma t} - 1)}\,. \label{eq:at}
\end{equation}
where $a_0$ is the constant of integration (the initial value of $a$, when $t=0$). The solution (\ref{eq:at}) only slightly differs from the standard inflationary solution, but for this solution we have $(\dot a)_{t=0} =0$.

Now we shall consider more realistic case. In this case we put the value of $\omega$ not equal to zero exactly, but to be very small, $\omega \approx 0, \quad \omega >0$. If we put now in Eq. (\ref{eq:bdot}) $\lambda_1 = \frac{3\lambda}{1+\omega}$, then this equation reads, 
\begin{eqnarray}
&& \dot v = -\frac{6\lambda}{1+\omega} \frac{e^{\omega v}-e^{-\omega v}}{2} \nonumber \\
&&= -\frac{6\lambda}{1+\omega} \sinh{\omega v} \sim -6\lambda \omega v\,, \label{eq:dotv}
\end{eqnarray}
with the solution
\begin{equation}
\beta = e^{Ce^{-6\lambda \omega t}} = \beta_0^{e^{-6\lambda \omega t}}\,, \label{eq:be}
\end{equation}
where $C$ is an integration constant, and $\beta_0 = e^C$ is a very large initial value of $\beta$, when $t=0$.

The solution (\ref{eq:be}) is more realistic then Eq. (\ref{eq:Bt}), because we have for this solution, 
\begin{eqnarray}
&&\beta \rightarrow 1\,, \quad \Lambda_{eff} = \beta^4 \Lambda \rightarrow \Lambda\,, \label{eq:brar}\\
&& {\rm when}\quad  t \rightarrow \infty\,.\nonumber
\end{eqnarray}
Therefore the limit of the effective cosmological constant for large time is not zero that coincides with cosmology observations and ensure  an accelerating expansion of modern universe. 

One can find the solution for the scale factor in explicit form only for some limiting cases, 
\begin{eqnarray}
&& a = a_0 e^{\lambda \beta_0 t}\,, \quad {\rm when} \quad t \quad {\rm is}\; {\rm small} \,, \label{eq:aebt}\\
&& a = a_0 e^{\lambda  t}\,, \quad {\rm when} \quad t \quad {\rm is}\; {\rm large} \,. \label{eq:aet}
\end{eqnarray}
As $\beta_0$ is very large, the universe inflation at the small times is more intensive, then for the standard inflation scenario.

This solution could be realized at the very beginning of the universe evolution, when the cosmological constant $\Lambda_0$ estimated by quantum field theory was equal $\Lambda_0 /\Lambda = \beta_0^4 \sim 10^{120}$, and the number $(\exp{2v})_{t=0} = \beta_0^2 \sim 10^{60}$ was very large.  

Our solutions are realized, if the following conditions are valid, 
\begin{equation}
\qquad B \approx 3A\,, \qquad B_1 = 1\,, \qquad B_2 = 1\,.  \label{eq:ABB}
\end{equation}
These 3 conditions in rather complicated manner is determined by the 16 coupling constants of the gravitational Lagrangian density (\ref{eq:LGG}), and can be easily fulfilled.   

We point out that the ultra-rapid decrease of the energy of physical vacuum according to the laws  (\ref{eq:Bt}) or (\ref{eq:be}) occurs only prior to the Friedman era evolution of the universe. Further evolution of the universe is determined not by a scalar field, but mainly by the born ultra relativistic matter and the radiation interacting with it. 

Thus our result can explain the exponential decrease in time at very early universe of the dark energy (the energy of physical vacuum), describing by the effective cosmological constant. This can give a way to solving the problem of cosmological constant as a consequence of fields dynamics at the early Universe. It is well-known that this problem is one of the fundamental problems of the modern fundamental physics \cite{Wein}.\\

This research work has been performed in the framework of the Federal Purposeful Program ``Research and Pedagogical Personnel of Innovative Russia for 2009-–2013''.

\end{document}